\documentclass[twocolumn,showpacs,amsmath,nofootinbib,pra,aps,amssymb,longbibliography]{revtex4-1}

\usepackage[T1]{fontenc}
\usepackage[utf8]{inputenc}
\usepackage{times}
\usepackage{color} 
\usepackage{array}
\usepackage{amssymb,amsmath}
\usepackage{amsbsy}
\usepackage[pdftex]{graphicx} 
\usepackage{bm}
\usepackage{float}
\usepackage{dcolumn}
 
\usepackage[unicode,breaklinks]{hyperref}
\hypersetup{
    unicode=true,
    plainpages=false, 
    colorlinks=true,
    linkcolor=blue,
    citecolor=blue,
    filecolor=black,
    urlcolor=blue
}
\usepackage{url}
\usepackage{verbatim}

\newcommand{\gQF}{g_{\mathrm{QF}}}

\newcommand{\Ut}{\tilde{U}}

\DeclareMathOperator{\Ei}{Ei}

\begin{document}

\title{Supersolidity in an elongated dipolar condensate}
	\author{P.~B.~Blakie,$^{1,2}$ D.~Baillie,$^{1,2}$ L.~Chomaz,$^{3}$ and F.~Ferlaino$^{3,4}$  }
	
	\affiliation{%
	$^1$Dodd-Walls Centre for Photonic and Quantum Technologies, New Zealand\\
	$^2$Department of Physics, University of Otago, Dunedin 9016, New Zealand\\
	$^{3}$Institut f\"ur Experimentalphysik, Universit\"at Innsbruck, Technikerstra{\ss}e 25, 6020 Innsbruck, Austria\\
	$^{4}$Institut f\"ur Quantenoptik und Quanteninformation, \"Osterreichische Akademie der Wissenschaften, Technikerstra{\ss}e 21a, 6020 Innsbruck, Austria}	 
\date{\today}
\begin{abstract}  
We present a theory for the emergence of a supersolid state in a cigar-shaped dipolar quantum Bose gas. 
Our approach is based on a reduced three-dimensional (3D) theory, where the condensate wavefunction is decomposed into an axial field and a transverse part described variationally.
 This provides an accurate fully 3D description that is specific to the regime of current experiments and efficient to compute. We apply this theory to understand the phase diagram for a gas in an infinite tube potential. We find that the supersolid transition has continuous and discontinuous regions as the averaged density varies. We develop two simplified analytic models to characterize the phase diagram and elucidate the roles of quantum droplets and of the roton excitation.    
\end{abstract} 

\maketitle

\begin{figure}[htbp!] 
   \centering
   \includegraphics[width=3.4in]{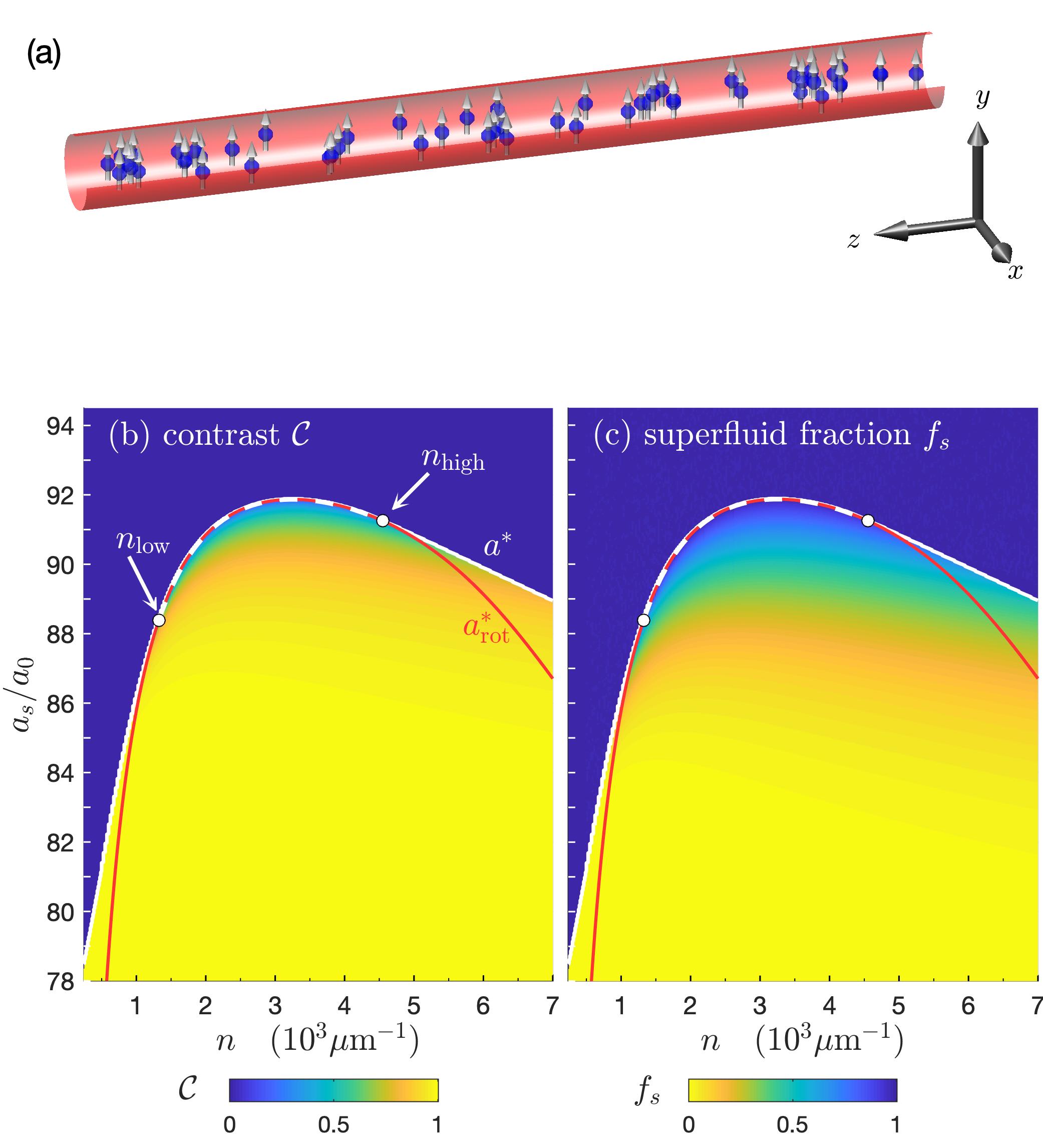}
   \caption{ (a) Schematic of the tube-dipolar system with confinement in the $xy$-plane and the dipoles oriented along the $y$-axis.  (b) Contrast of density modulation and (c) superfluid fraction as a function of the average linear density and the s-wave scattering length.  The s-wave scattering length for the transition to the modulated state ($a^*$, white line) and for the roton instability of the uniform BEC ($a^*_{\text{rot}}$, red line)  are shown. The dashed line (where $a^*=a^*_\text{rot}$) between the circle markers at the densities $n_\text{low}=1.3\times10^3/\mu$m and $n_\text{high}=4.6\times10^3/\mu$m indicates a continuous transition.
Results for  $^{164}$Dy using $a_{dd}=130.8a_0$, with $\omega_{x,y}=2\pi\times150$Hz.
   }
   \label{fig:schematicPD}
\end{figure}

\noindent\textbf{Introduction} -
A supersolid is a state of matter exhibiting both crystalline order and superfluidity  \cite{Gross1957a,Andreev1969a,Chester1970a,Leggett1970a,Boninsegni2012a}. 
Studies of solid $^4$He have yet to yield a clear signature of supersolidity \cite{Chan2004a,Kim2012a,Boninsegni2012a}, and many efforts have turned to dilute ultra-cold atomic systems \cite{Henkel2010a,Saccani2012a,Cinti2010a,Macri2013a,Ancilotto2013a,Lu2015a,Leonard2017a,Li2017a,Wenzel2017a,Baillie2018a,Roccuzzo2019a,Zhang2019a}. Recently three experiments have reported the observation of a supersolid in a dipolar Bose-Einstein condensate (BEC) \cite{Tanzi2019a,Bottcher2019a,Chomaz2019a}, and have studied its elementary excitations \cite{Tanzi2019b,Guo2019a,Natale2019a}. These experiments have all used an elongated trap with the atomic magnetic dipoles polarized along a tightly confined  direction [e.g.~see Fig.~\ref{fig:schematicPD}(a)]. 
The supersolid transition was explored by reducing the s-wave scattering length below a critical value  whereby crystalline order (spatial density modulation of the gas) develops along the weakly confined direction.

Calculations of the ground states of this system using the extended Gross-Pitaevskii equation (eGPE) have shown good quantitative agreement with the observations of the experiments. The eGPE theory differs from the usual Gross-Pitaevskii equation for dipolar condensates by including the leading order effect of quantum fluctuations \cite{Ferrier-Barbut2016a,Chomaz2016a,Wachtler2016a,Bisset2016a}. Currently little is known about the phase diagram, such as the nature of the transition between phases and how this depends on the confining potential or atomic density. Previous general studies of supersolidity have predicted that the transition is continuous in the one-dimensional (1D) case, while higher dimensional cases are generally discontinuous \cite{Sepulveda2008a} (also see \cite{Lu2015a}). However, while the supersolid realized in dipolar gas experiments exhibits crystalline order (density modulation) along one spatial dimension, the transverse degrees of freedom are not frozen out. This intrinsic 3D character is not captured in earlier theories of 1D supersolids, and developing an appropriate theory for this regime is the focus of this paper.

Here we develop formalism for a dipolar gas in an infinite tube, i.e.~with transverse harmonic confinement but free in the $z$-direction. We develop a simplified 3D theory in which the transverse wavefunction is described by two variational parameters and the axial field is treated numerically (cf.~the planar theory of  Ref.~\cite{Zhang2019a}).
 Our main results are the phase diagrams in  Fig.~\ref{fig:schematicPD}(b) and (c) revealing the contrast of the density modulations (i.e.~crystalline order) and the persistence of superfluidity, respectively. These results show that the condensate to supersolid transition is continuous for a range of intermediate densities, but is otherwise discontinuous.

\noindent\textbf{Reduced 3D theory for the elongated system} -
We consider a zero temperature dipolar Bose gas described by the field $\Psi(\mathbf{x})$. The transverse confinement is harmonic with angular frequencies ($\omega_x,\omega_y$) and the atomic magnetic dipoles are aligned along the $y$-direction by an external magnetic field [see Fig.~\ref{fig:schematicPD}(a)]. Following \cite{Blakie2020a} we  decompose the field as $\Psi(\mathbf{x})=\psi(z)\chi(x,y)$ where $\psi(z)$ describes the axial field and $\chi$ is a variational treatment of the transverse directions. We take  $\chi(x,y)=\tfrac{1}{\sqrt{\pi}l} {e^{-(\eta x^2+y^2/\eta)/2l^2}}$ to be a Gaussian function with variational parameters $\{l,\eta\}$ describing its mean width and anisotropy.
Our interest is in stationary solutions of specified average linear density $n$ along $z$.

The uniform ground state is the form $\Psi_\text{BEC}=\sqrt{n}\chi$, which we refer to as the BEC state. Here the energy per particle, computed from the eGPE energy functional, is given by the nonlinear function \begin{align}
\mathcal{E}_\text{BEC}(l,\eta) =\mathcal{E}_\perp+ \frac{1}{2}n \Ut(0)+\frac{2}{5}g_{\mathrm{QF}}n^{\frac{3}{2}},\label{Eq:Eu}
 \end{align}
 where $\mathcal{E}_\perp = \frac{\hbar^2}{4ml^2}(\eta+\frac{1}{\eta})+\frac{ml^2}{4}(\frac{\omega_x^2}{\eta}+\omega_y^2\eta)$ is the single-particle energy of the transverse degrees of freedom. Note that $\tilde{U}$ and $g_{\mathrm{QF}}$ also depend on $l$ and $\eta$ \cite{Lima2011a,Ferrier-Barbut2016a,Chomaz2016a,Wachtler2016a,Bisset2016a}.
The two-body interaction for the axial field in $k_z$-space is 
 \begin{align}
\!\!\Ut(k_z)&\!=\!\frac{2a_s\hbar^2}{m l^2}\! +\!\frac{2a_{dd}\hbar^2}{m l^2}\!\left\{\!\frac{3[ Qe^{Q}\Ei(-Q)+1]}{1+\eta}-1\!\right\},\label{Ueta} 
\end{align} 
with  $\Ei$ being the exponential integral, and $Q\equiv \tfrac{1}{2}\sqrt{\eta}k_z^2l^2$ \cite{Blakie2020a}. In Eq.~(\ref{Ueta}) $a_s$ is the $s$-wave scattering length and $a_{dd}\equiv m\mu_0\mu_m^2/12\pi\hbar^2$ is the dipole length. The quantum fluctuations are described in Eq.~(\ref{Eq:Eu}) by the higher order nonlinearity with  coefficient $g_{\mathrm{QF}}=\frac{256\hbar^2}{15 ml^3}a_s\sqrt{ a_s^3}(1+\tfrac32 \epsilon_{dd}^2)$, 
where $\epsilon_{dd}=a_{dd}/a_s$. 
 In  Ref.~\cite{Blakie2020a} the accuracy of this variational theory has been established with detailed comparisons to full numerical solutions of the 3D eGPE.  

Of most interest here is when the ground state of the system spontaneously breaks the translational symmetry along $z$ and develops crystalline order, i.e.~$|\psi|^2$ is periodically modulated in space. In this case we can define  $\psi$ on a unit cell of length $L$, i.e.~$\mathrm{uc}=\{-\frac{1}{2}L\le z<\frac{1}{2}L\}$, subject to periodic boundary conditions, and the normalization constraint  $\int_{\mathrm{uc}} dz\,|\psi|^2= nL$. 
The energy per particle is given by
\begin{align}
\!\mathcal{E}\!=\!\mathcal{E}_\perp\!+\!\int_{\mathrm{uc}}\!\frac{dz}{ nL}\,\psi^*
\!\left(\!-\frac{\hbar^2}{2m}\frac{d^2}{dz^2} +\frac{1}{2}\Phi +\frac{2}{5}g_{\mathrm{QF}}|\psi|^3\right)\!\psi\label{eGPE_Einf}, 
\end{align}
where the two-body interactions are described by the effective potential  $\Phi(z)=\mathcal{F}_z^{-1}\left\{\Ut(k_z )\mathcal{F}_z\{|\psi|^2\}\right\}$,  with $\mathcal{F}_z$ being the 1D Fourier transform. To obtain stationary solutions we vary $\{l,\eta,L\}$ and $\psi(z)$ to find local minima of (\ref{eGPE_Einf}).

\noindent\textbf{Phase diagram} - In Fig.~\ref{fig:schematicPD}(b) and (c) we present phase diagrams for the development of crystalline and superfluid order found by solving the reduced 3D eGPE for the ground state by minimising Eq.~(\ref{eGPE_Einf}) as a function of $n$ and $a_s$. Our results are for Dy atoms in a radially isotropic tube potential with $\omega_{x,y}=2\pi\!\times\!150$Hz, similar to the transverse trapping used in experiments \cite{Tanzi2019a, Bottcher2019a,Chomaz2019a,Tanzi2019b,Natale2019a,Guo2019a}. 

\begin{figure}[htbp] 
   \centering
   \includegraphics[width=3.4in]{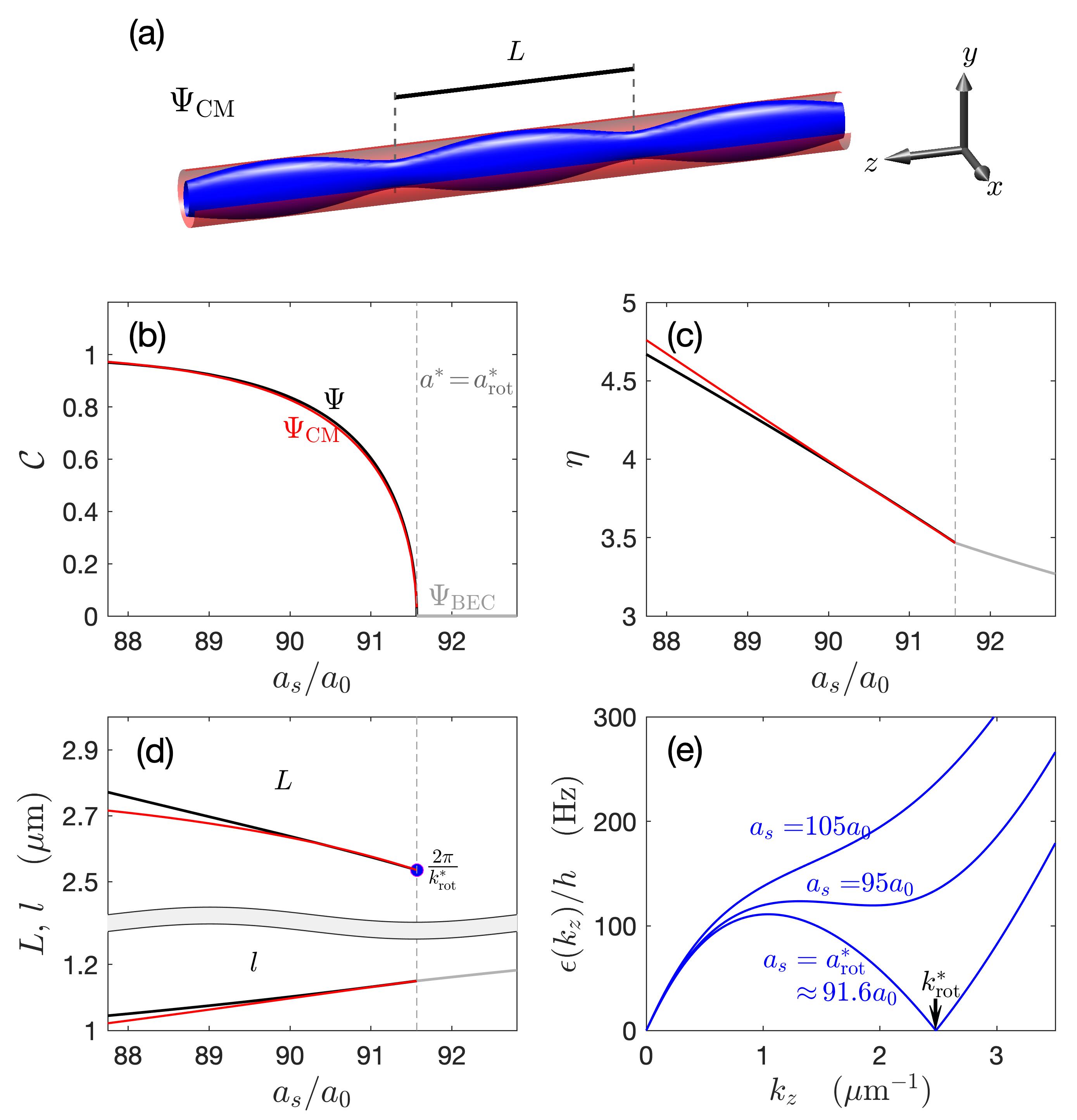}
   \caption{Continuous transition to density modulated state. (a) Schematic  showing a density isosurface of $|\Psi_\text{CM}|^2$.  
   (b) Contrast, (c) transverse anisotropy $\eta$, and the (d) transverse width $l$ and unit cell size $L$, as $a_s$ varies.  In (b)-(d) we compare the results of the full theory for $\Psi$  (black line) to the model  $\Psi_\text{CM}$ (red line). In the uniform BEC state both theories are identical (grey line). The vertical dashed line indicates the transition point $a^*$. In (d) the length  $L$ is only uniquely defined in the modulated state and the critical roton wavelength $2\pi/k_\text{rot}^*$ [see (e)] is shown for reference. (e) Excitation spectrum of uniform BEC state.  
   Parameters as in Fig.~\ref{fig:schematicPD} with $n=2.5\times10^3/\mu$m. }
   \label{fig:ctstransition}
\end{figure}

\begin{figure}[htbp] 
   \centering
   \includegraphics[width=3.4in]{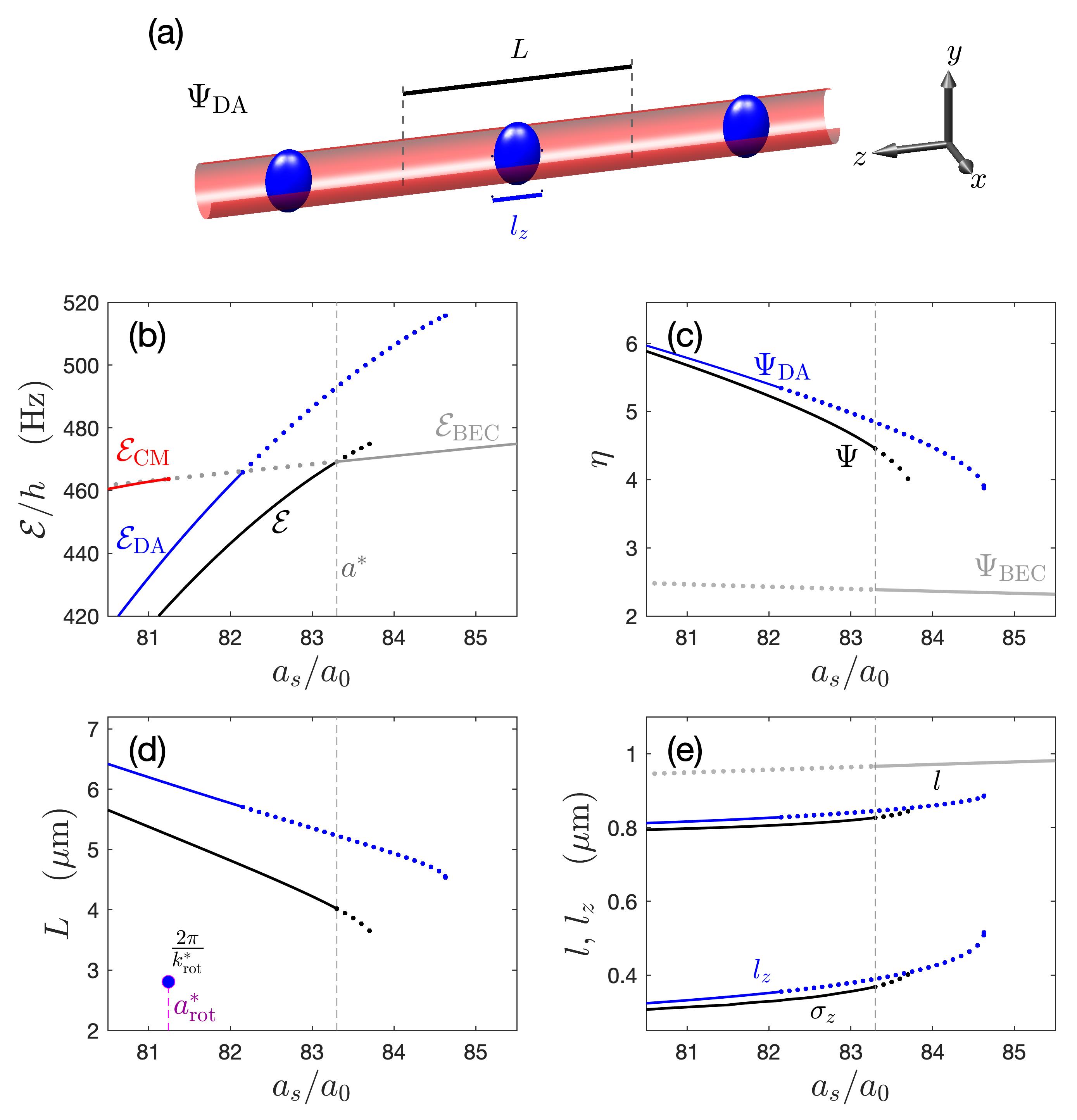}  
   \caption{Low-density discontinuous transition. (a) Schematic showing a density isosurface of $|\Psi_\text{DA}|^2$.   Comparison of  $\Psi$ (black line/black dotted line) and $\Psi_\text{DA}$ (blue line/blue dotted line) results for the (b) energy per particle, (c) transverse anisotropy $\eta$, and   (d,e) length scales as $a_s$ varies. Solid (dotted) lines are used for each theory when its energy is lower (higher) than $\Psi_\text{BEC}$. In (b) the CM ansatz results are also shown (red line).
   In (d)  the critical roton wavelength is  indicated at $a^*_\text{rot}$. In (e) $\sigma_z$ is the $1/e$ halfwidth of $|\Psi|^2$  along the $z$ axis.   
   Parameters as in Fig.~\ref{fig:schematicPD} with $n=0.7\times10^3/\mu$m.}
   \label{fig:lowDtransition}
\end{figure}

\begin{figure}[htbp] 
   \centering
   \includegraphics[width=3.4in]{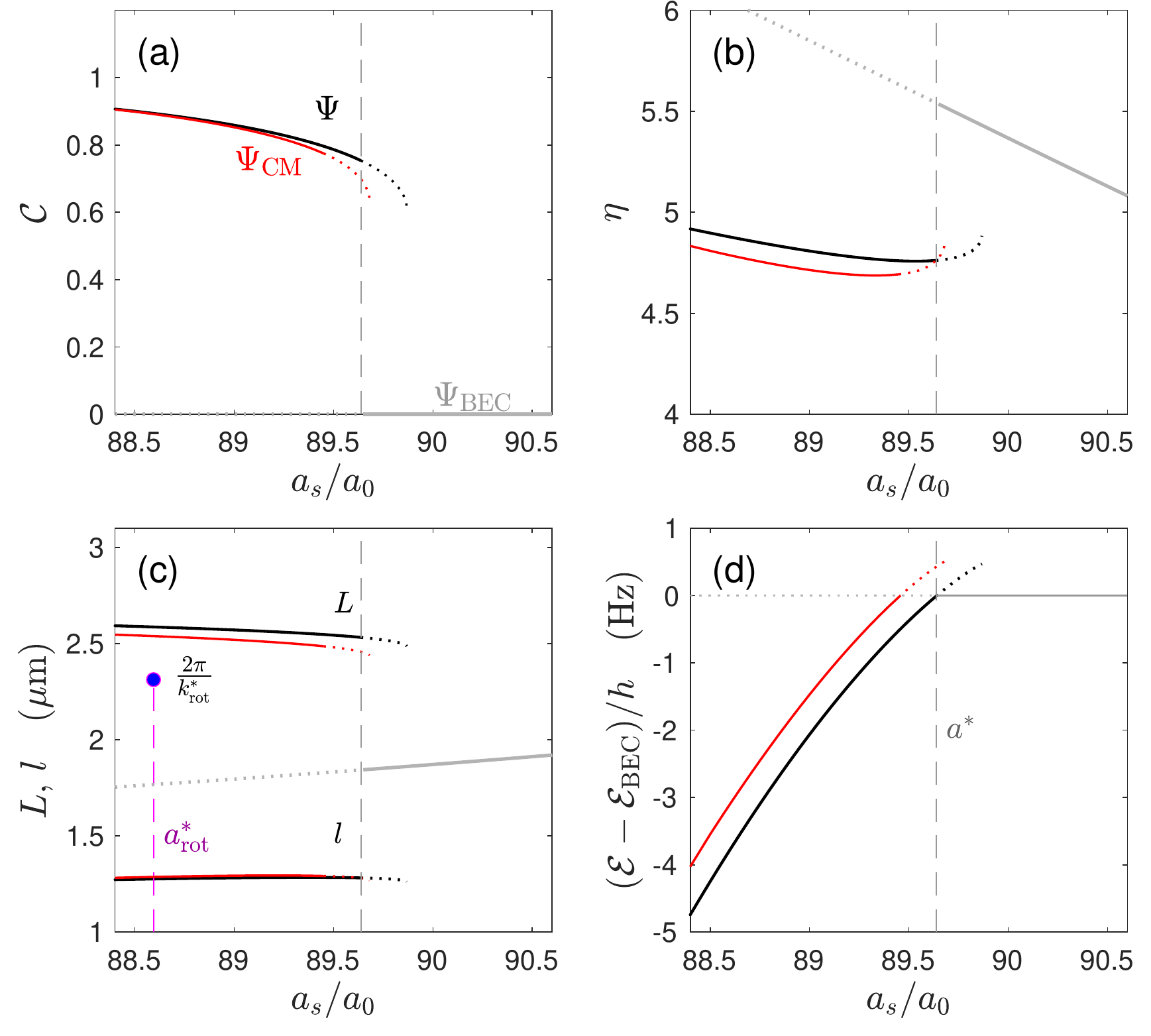}
   \caption{High-density discontinuous transition. 
   Comparison of  $\Psi$ (black line/black dotted line) and $\Psi_\text{CM}$ (red line/red dotted line) results for the (a) contrast, (b) transverse anisotropy $\eta$, and the (c) length scales $l$ and $L$ as $a_s$ varies. In (c) we also indicate the roton wavelength at $a_\text{rot}^*$ for reference. (d) Energy per particle comparison of the theories. The solid lines are used for each theory when its energy is lower than $\Psi_\text{BEC}$. 
     Parameters as in Fig.~\ref{fig:schematicPD} with $n=6.25\times10^3/\mu$m. }
   \label{fig:highDtransition}
\end{figure}

We characterize the crystalline order by the linear density contrast   
$\mathcal{C}=(|\Psi|^2_{\max}-|\Psi|^2_{\min}|)/(|\Psi|^2_{\max}+|\Psi|^2_{\min}),$ where $|\Psi|_{\max}$ and $|\Psi|_{\min}$ are the maximum and minimum of $|\Psi|$ on the $z$ axis, respectively.
When $\mathcal{C}=0$ the ground state is uniform and is identical to $\Psi_\text{BEC}$.
The results in Fig.~\ref{fig:schematicPD}(b)  show that density-modulated ground states occur below a certain scattering length $a^*$ that depends on $n$.  For an intermediate range of densities $n\in[n_\text{low},n_\text{high}]$ the transition between uniform and modulated states is continuous, i.e.~the contrast develops continuously and the ground state wavefunction evolves smoothly as $a_s$ changes [also see Fig.~\ref{fig:ctstransition}(b)-(d)]. Outside of this density range the transition is discontinuous, i.e.~with a discontinuity in contrast and a sudden change in the ground state wavefunction (also see Figs.~\ref{fig:lowDtransition} and \ref{fig:highDtransition}).

We also quantify the superfluid order in the system using the bound proposed by Leggett \cite{Leggett1970a,Sepulveda2008a}
\begin{align}
f_s=\frac{L}{n}\left\{\int_{\mathrm{uc}} \frac{dz}{|\psi|^2}\right\}^{-1}.\label{Eq:fs}
\end{align}
For states that modulate along one spatial dimension this measure is equivalent to the superfluid fraction based on the nonclassical translation inertia \cite{Roccuzzo2019a,Sepulveda2010a}. The results in Fig.~\ref{fig:schematicPD}(c) show that the BEC has $f_s=1$. This reduces when the density modulates, but remains appreciable close to the transition line $a^*$ and for $n\gtrsim1\times10^3/\mu$m. We regard the modulated ground state with an appreciable superfluid fraction as being in the supersolid phase.   As the scattering length is reduced the superfluid fraction rapidly reduces and the system becomes an insulating droplet crystal without any appreciable superfluid transport.  In this regime quantum and thermal fluctuations (not included in the current theory) will be important and may cause the crystal to insulate at higher $a_s$ values.
 
\noindent\textbf{Continuous transition region} - Weakly density-modulated states are well-described by the cosine-modulated (CM) ansatz,  $\Psi_\text{CM}(\mathbf{x})=\psi_\text{CM}(z)\chi(x,y)$,  where
\begin{align}
\psi_\text{CM}(z)=\sqrt{n} (\cos\theta+\sqrt{2}\sin\theta\cos \tfrac{2\pi z}{L} ),\label{psivar}
\end{align}
(also see \cite{Sepulveda2008a,Lu2015a,Zhang2019a}).  The CM ansatz replaces the axial wavefunction by the variational parameters $\theta$, describing the amplitude of the density modulation, and $L$, specifying the wavelength of the modulation 
[see Fig.~\ref{fig:ctstransition}(a)]. Thus $\Psi_\text{CM}$ depends on the four variational parameters
$\{\theta,L,l ,\eta\}$.
Note that the factor of $\sqrt{2}$ ensures that this ansatz has an average density $n$ per unit cell (independent of $\theta$).
We restrict $\theta$   to the range\footnote{For $\theta>\varphi$,   $|\psi_\text{CM}|^2$ has two local maxima per unit cell, which we regard as an artefact. All CM analytic results presented here are for $0\le \theta\le\varphi$.} $\theta\in[0,\varphi]$, where $\varphi\equiv \cot^{-1} \sqrt2\approx0.616$.    We note that $\theta$ directly relates to the density contrast  as $\mathcal{C}(\theta)=  \frac{2\sqrt2 \sin 2\theta}{3-\cos 2\theta}$, with $\mathcal{C}(\varphi)=1$.  The superfluid fraction (\ref{Eq:fs}) is $f_s(\theta)=\cos^2\theta(1-2\tan^2\theta)^{3/2}$, which decreases with increasing $\theta$, until  $f_s(\varphi)=0$.
Using this ansatz we can analytically evaluate the energy per particle as   \begin{align}
\mathcal{E}_\text{CM}(\theta,L,l ,\eta)&=\mathcal{E}_\perp+\frac{h^2\sin^2\!\theta}{2mL^2}+ \frac{2}{5}g_{\mathrm{QF}}n^{3/2}\Lambda(\theta)\label{var_ECM}\\
+ \frac{n}{2} &\left[\Ut(0) + \sin^2\!2\theta \Ut(\tfrac{2\pi}{L})   +\tfrac{1}{2}\sin^4\!\theta \Ut(\tfrac{4\pi}{L})   \right]\nonumber,
\end{align} 
where 
 $\Lambda(\theta) =   \frac1{32}(90 \cos \theta - 55 \cos 3\theta - 3 \cos 5\theta)$. 
We determine solutions  $\Psi_\text{CM}$  by numerically minimising $\mathcal{E}_\text{CM}$ with respect to its variational parameters. We expect this ansatz to accurately describe the system's behavior in the vicinity of the continuous transition.

In Fig.~\ref{fig:ctstransition}(b)-(d) we present results for the ground state properties at an intermediate density of $n=2.5\times10^3/\mu$m. 
We compare the properties of the ground states: $\Psi$ obtained from the reduced 3D model  [minimising Eq.~(\ref{eGPE_Einf})] and $\Psi_\text{CM}$   [minimising Eq.~(\ref{var_ECM})].
Using the reduced 3D theory we find that the system undergoes a continuous transition from the uniform BEC state into a modulated state at $a^*\approx91.6a_0$.  Both theories are in excellent agreement close to the transition and show that for $a_s<a^*$ the contrast  initially develops as $\mathcal{C}\sim\sqrt{a^*-a_s}$. For $a_s$ values further below $a^*$, the theories begin to deviate as a strong density modulation develops.   

Our results also show the importance of the full 3D description of the system [see Figs.~\ref{fig:ctstransition}(c) and (d)]. While the confinement is radially isotropic magnetostriction causes $\chi$ to highly elongate in the dipole direction with an anisotropy of $\eta\sim4$. Interactions also cause $l$ to be significantly greater than the harmonic oscillator length $l_\text{ho}=0.64\mu$m.

Using the  $\Psi_\text{CM}$ ansatz we can identify the conditions for the continuous transition by looking for stationary points of ${\mathcal{E}}_\text{CM}$ for small $\theta$.  To leading order in $\theta$ we have $\frac{\partial {\mathcal{E}_\text{CM}}}{\partial\theta}=\theta\Delta(\tfrac{2\pi}{L})$, where  
\begin{align}
\Delta(k)= \epsilon_0(k)+2n\Ut(k)+3\gQF n^{3/2},
\end{align}
with $\epsilon_0\equiv\hbar^2k^2/2m$.
Thus the stationary points of $\mathcal{E}_\text{CM}$ in this regime are either: (i) a uniform solution with $\theta=0$  (in which case $L$ is irrelevant); (ii) the modulated solutions with $\theta\ne0$ and $L$ determined by   $\Delta(\tfrac{2\pi}{L})=0$.
We note that the Bogoliubov spectrum for the excitations of the uniform condensate $\theta=0$ is given by 
$\epsilon(k)=\sqrt{\epsilon_0(k)\Delta(k)}$  \cite{Blakie2020a}, thus the condition $\Delta(\tfrac{2\pi}{L})=0$  means that an excitation of wavelength $L$ has zero energy, i.e.~a roton-like excitation in the system goes soft \cite{Santos2003a,Chomaz2018a}.  In Fig.~\ref{fig:ctstransition}(e) we show the uniform BEC excitation spectrum for various $a_s$ values, observing the formation of a roton at $a_s\approx95a_0$ that softens to zero energy at $91.6a_0$. This marks the dynamic instability of the uniform BEC state and defines the roton critical value $a^*_\text{rot}$.
At this critical point the modulated state develops with a wavelength corresponding to the roton wavevector
 [Fig.~\ref{fig:ctstransition}(d)].
  In the regime where the transition is continuous  $a^*$ coincides with $a^*_\text{rot}$. We see that this holds for the results in  Figs.~\ref{fig:schematicPD}(b) and (c), with  $a^*_\text{rot}=a^*$ for $n\in[n_\text{low},n_\text{high}]$. Outside of this density range (where the transition is discontinuous) we see that $a^*_\text{rot}<a^*$.

\noindent\textbf{Low-density discontinuous  transition region} - For densities $n<n_\text{low}$ the contrast has a finite jump as the transition to the modulated state is crossed. In Figs.~\ref{fig:lowDtransition}(b)-(e) we focus in on this regime of the phase diagrams in Fig.~\ref{fig:schematicPD}.
We consider the particular case of $n=0.7\!\times\!10^3/\mu$m with a transition point at $a^*=83.3a_0$, identified as where the energy per particle $\mathcal{E}$ of the modulated stationary solution $\Psi$ crosses $\mathcal{E}_\text{BEC}$ [Fig.~\ref{fig:lowDtransition}(b)]. This transition point is appreciably higher than $a^*_\text{rot}=81.2a_0$, where the BEC becomes dynamically unstable, meaning that hysteresis can occur in $a_s$ ramps across the transition with the uniform or modulated state persisting as a metastable state. Results of the reduced 3D theory shows that the transition occurs with a sudden jump in the transverse properties of the ground state [Figs.~\ref{fig:lowDtransition}(c) and (e)], and that the unit cell size is larger than (and disconnected from) the roton wavelength [Fig.~\ref{fig:lowDtransition}(d)]. The contrast (not shown) remains close to unity up until the modulated solution branch terminates as a metastable state.

The $\Psi_\text{CM}$ ansatz fails to describe  this regime [e.g.~Fig.~\ref{fig:lowDtransition}(b)] and incorrectly predicts the transition to occur continuously at the point of roton softening. Here the reduced 3D model indicates that the system prefers to organize into localized well separated droplets, exhibiting a high contrast modulation of the density. This motivates us to introduce the droplet array ansatz $\Psi_\text{DA}(\mathbf{x})=\psi_\text{DA}(z)\chi(x,y)$, where
\begin{align}
 \psi_\text{DA}(z)= \sqrt{ \tfrac{N_\text{D}}{\sqrt{\pi}l_z}} {  {e^{-z^2/2l_z^2}}},\label{Eq;DQ}
 \end{align}
represents a Gaussian droplet of $z$-width $l_z$  and containing $N_\text{D}$ atoms. The droplets repeat every $L$ [i.e.~one per unit cell,  see Fig.~\ref{fig:lowDtransition}(a)], with the relation $N_\text{D}=nL$ ensuring that the average density is fixed to $n$. 
We require well separated droplets ($L\gg l_z$) for  $\psi_\text{DA}$ to avoid a discontinuity at the unit cell boundary.
 We can evaluate the energy per particle of this ansatz   
 \begin{align}
  \mathcal{E}_\text{DA} =&\mathcal{E}_\perp+\frac{\hbar^2}{4ml_z^2}+\frac{\hbar^2nL}{\sqrt{2\pi} ml ^2l_z}\left(a_s+a_{dd}\frac{f(\kappa)+1-\eta}{1+\eta}\right) 
\nonumber \\ 
 &+ {\gQF}\left(\frac{2nL}{5\sqrt{\pi}l_z}\right)^{3/2} +
\frac{3na_{dd}\hbar^2}{mL^2}\zeta(3),\label{Ed}
 \end{align}
 where $\kappa=\eta^{1/4}l /l_z$, $\zeta$ is the Riemann zeta function,  and 
$ f=\tfrac{1 + 2\kappa^2 - 3\kappa^2\mathrm{atanh}\sqrt{1 - \kappa^2}/\sqrt{1 - \kappa^2}}{1 -\kappa^2}$
   (also see Ref.~\cite{Lima2010a}). 
All terms in Eq.~(\ref{Ed}) are evaluated exactly except the last term describing the long-range interaction between droplets. That term is obtained by approximating each droplet as a point dipole of $N_\text{D}$ atoms, which is a good approximation for our regime with $l_z\ll L$.  The  $\Psi_\text{DA}$ solutions are determined by numerically minimising $  \mathcal{E}_\text{DA}$ with respect to $\{L,l_z,l,\eta\}$.

The results in Figs.~\ref{fig:lowDtransition}(b)-(e) show that $\Psi_\text{DA}$ is in good agreement with $\Psi$. In Fig.~\ref{fig:lowDtransition}(e) we show the droplet width $\sigma_z$ ($l_z$) along $z$ from $\Psi$ ($\Psi_\text{DA}$), and observe that these remain  much smaller than their spacing $L$ [Fig.~\ref{fig:lowDtransition}(d)] up until their solution branch terminates.
At the termination point the modulated solution is metastable ($\mathcal{E}_\text{BEC}$  is lower), and occurs because the droplets unbind ($l_z,\sigma_z\to \infty$).

\noindent\textbf{High-density discontinuous  transition region} -  
In Figs.~\ref{fig:highDtransition}(a)-(d) we focus in on the high-density ($n>n_\text{high}$) discontinuous transition of  the phase diagrams in Fig.~\ref{fig:schematicPD}. We consider the particular case of  $n=6.25\times10^3/\mu$m, where the transition occurs at $a^*=89.6a_0$ (cf.~$a_\text{rot}^*=88.6a_0$).  As we cross the transition the ground state $\Psi$ suddenly changes its transverse profile ($\eta$ and $l$) and develops a strong density modulation. In this regime the discontinuous transition arises from an interplay between the transverse degrees of freedom and the modulation along $z$ mediated by the dominant role of interactions at higher densities.  
The behavior of $\Psi$ is well described by the $\Psi_\text{CM}$ ansatz, however the small-$\theta$ stationary point of $\Psi_\text{CM}$ that we analyzed earlier to understand the continuous transition is an unstable saddle point in this regime. 
Here the (meta)-stable CM solution found   [see Figs.~\ref{fig:highDtransition}(a)-(d)] has a large $\theta$ value (i.e.~high contrast)  and a longer wavelength than the roton wavelength [Fig.~\ref{fig:highDtransition}(c)]. This solution also predicts transverse properties $(l,\eta)$ and $a^*>a_\text{rot}^*$, similar to the reduced 3D theory. We also note that at this density the DA solution terminates at $a_s\sim80a_0$, well below the scattering length range considered in Fig.~\ref{fig:highDtransition}.

\begin{figure}[htbp] 
   \centering
   \includegraphics[width=3.2in]{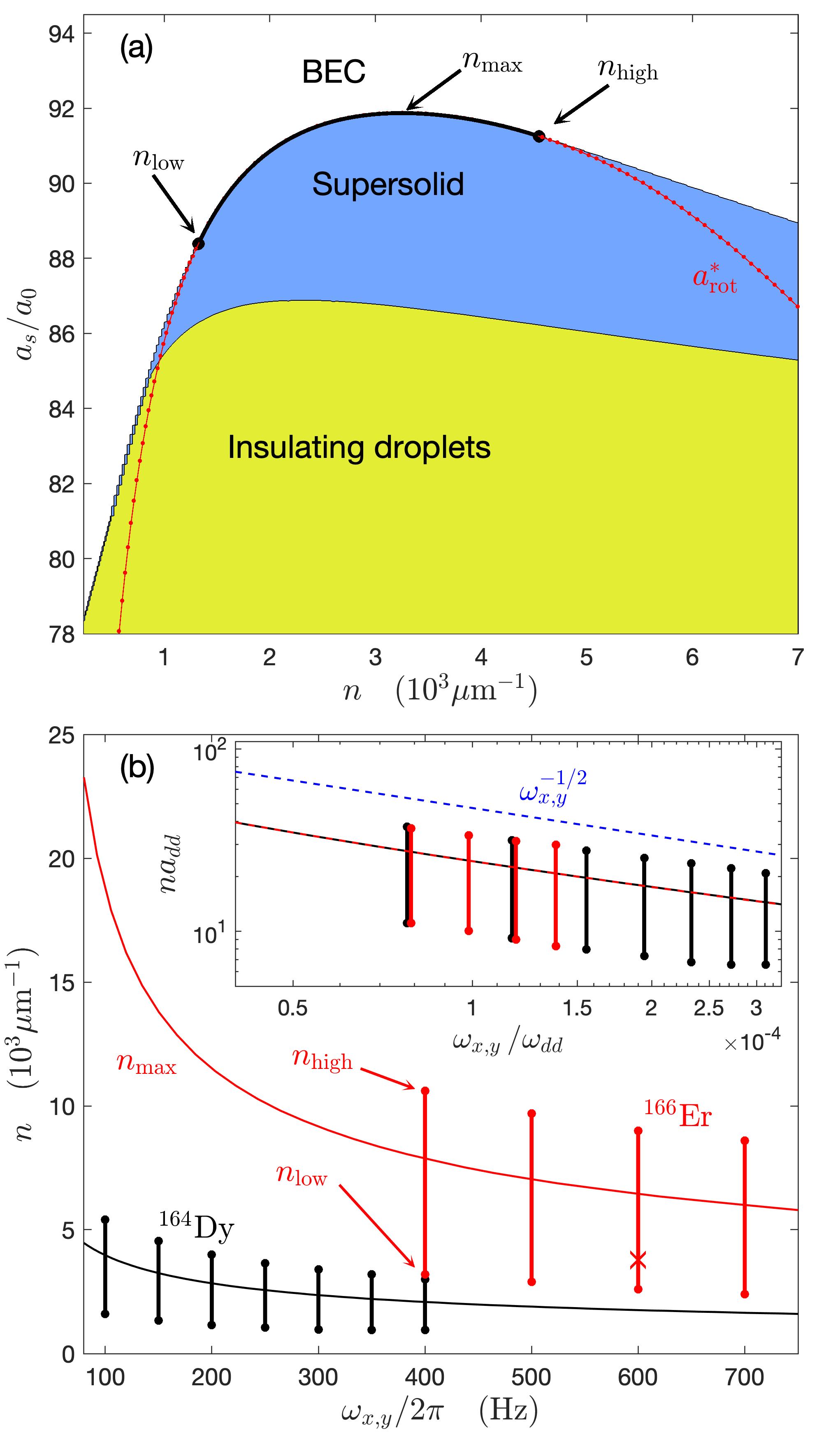}
   \caption{(a) Phase diagram summarizing the results of Fig.~\ref{fig:schematicPD} with $n_{\max}=3.25\times10^3/\mu$m being where $a^*_\text{rot}$ is maximized. (b) The value of $n_{\max}$ versus  $\omega_{x,y}$ for $^{166}$Er and $^{164}$Dy systems. The vertical bars show the continuous transition range $[n_\text{low},n_\text{high}]$ for several cases. The cross indicates the parameters of the calculation in Ref.~\cite{Roccuzzo2019a}. The inset shows the rescaled data.}
   \label{fig:schematicPDX}
\end{figure}

\noindent\textbf{Summary and outlook} - 
We have developed theory for a tube confined dipolar quantum gas to understand its phase diagram. In Fig.~\ref{fig:schematicPDX}(a) we summarize the phase diagram quantitatively presented in Fig.~\ref{fig:schematicPD}, indicating the three phases. At low densities there is a direct discontinuous transition between the insulating droplet\footnote{Here we take the insulating droplet phase as a modulated state with $f_s\le0.1$ (an arbitrarily chosen  small superfluid fraction).}  and BEC phases. At higher densities a supersolid phase emerges, separating the BEC and insulating droplet phases, and the BEC-supersolid transition can be continuous within a certain density range.

In Fig.~\ref{fig:schematicPDX}(b) we characterize how the phase diagram changes with system parameters. We show the continuous transition range $[n_\text{low},n_\text{high}]$ as vertical lines obtained from phase diagrams like Fig.~\ref{fig:schematicPDX}(a) computed for the two relevant experimental atomic species (Er and Dy) and various (isotropic) transverse confinement strengths. 
We also show  $n_{\max}$, defined as the density where $a_\text{rot}^*$ (and $a^*$) is maximised.
 These results indicate that the characteristic densities $\{n_\text{low},n_\text{high},n_{\max}\}$ decrease with increasing radial trapping. Also that a continuous transition in Er requires a higher density than Dy, which will shorten the system lifetime due to three-body loss. A recent   3D eGPE calculation by Roccuzzo \textit{et al.}~\cite{Roccuzzo2019a} was performed for the tube-dipolar system and found a small jump in $f_s$ in the BEC-supersolid transition [for parameter set marked "x" in Fig.~\ref{fig:schematicPDX}(b)]. This may indicate the weak first order transition, and that the continuous region is narrower in  the full theory.
 
By rescaling the eGPE using $a_{dd}$  and $\omega_{dd}=\hbar/ma_{dd}^2$ as units of length and frequency, respectively, the resulting equation only depends on the scaled axial density $na_{dd}$, trap frequencies $\omega_{x,y}/\omega_{dd}$, and $\epsilon_{dd}$. The inset to Fig.~\ref{fig:schematicPDX}(b) shows the collapse of the rescaled Er and Dy  results in these units. We also observe that the characteristic densities scale with the transverse confinement frequency as $\omega_{x,y}^{-1/2}$.

The phase diagram and scaling we have predicted could be explored in future experiments, e.g.~by changing the atom number to pass through the transition at different densities, or by changing the transverse confinement to shift the location of the continuous transition region.  Additionally, we note that having tighter confinement along the dipole direction relative to other transverse direction (i.e.~$\omega_y>\omega_x$) can introduce additional discontinuous transitions into the phase diagram, similar to those observed in oblate pancake shaped traps \cite{Blakie2016a,Ferrier-Barbut2018a}.  The CM and DA analytic models we have presented provide simple tools to map out the phase diagrams and should aid in exploring and better understanding this fascinating system. Another important future step will be to include the axial trapping potential which will introduce finite size effects and a spatially dependent mean axial density.
\vspace*{-0.5cm}\begin{acknowledgments}
PBB and DB acknowledge the contribution of NZ eScience Infrastructure (NeSI) high-performance computing facilities, and support from the Marsden Fund of the Royal Society of New Zealand. FF and LC acknowledge the support from the European Commission via an ERC Consolidator Grant (RARE, no.\,681432), from the Austrian Science Fund (FWF) via a joint FWF/DPG FOR grant (FOR 2247/PI2790) and a joint FWF/RSF grant (I 4426). LC also acknowledges the support from the FWF via an Elise Richter Fellowship (V792).
 
\vspace{5mm}
\end{acknowledgments}


\begin{thebibliography}{38}%
\makeatletter
\providecommand \@ifxundefined [1]{%
 \@ifx{#1\undefined}
}%
\providecommand \@ifnum [1]{%
 \ifnum #1\expandafter \@firstoftwo
 \else \expandafter \@secondoftwo
 \fi
}%
\providecommand \@ifx [1]{%
 \ifx #1\expandafter \@firstoftwo
 \else \expandafter \@secondoftwo
 \fi
}%
\providecommand \natexlab [1]{#1}%
\providecommand \enquote  [1]{``#1''}%
\providecommand \bibnamefont  [1]{#1}%
\providecommand \bibfnamefont [1]{#1}%
\providecommand \citenamefont [1]{#1}%
\providecommand \href@noop [0]{\@secondoftwo}%
\providecommand \href [0]{\begingroup \@sanitize@url \@href}%
\providecommand \@href[1]{\@@startlink{#1}\@@href}%
\providecommand \@@href[1]{\endgroup#1\@@endlink}%
\providecommand \@sanitize@url [0]{\catcode `\\12\catcode `\$12\catcode
  `\&12\catcode `\#12\catcode `\^12\catcode `\_12\catcode `\%12\relax}%
\providecommand \@@startlink[1]{}%
\providecommand \@@endlink[0]{}%
\providecommand \url  [0]{\begingroup\@sanitize@url \@url }%
\providecommand \@url [1]{\endgroup\@href {#1}{\urlprefix }}%
\providecommand \urlprefix  [0]{URL }%
\providecommand \Eprint [0]{\href }%
\providecommand \doibase [0]{http://dx.doi.org/}%
\providecommand \selectlanguage [0]{\@gobble}%
\providecommand \bibinfo  [0]{\@secondoftwo}%
\providecommand \bibfield  [0]{\@secondoftwo}%
\providecommand \translation [1]{[#1]}%
\providecommand \BibitemOpen [0]{}%
\providecommand \bibitemStop [0]{}%
\providecommand \bibitemNoStop [0]{.\EOS\space}%
\providecommand \EOS [0]{\spacefactor3000\relax}%
\providecommand \BibitemShut  [1]{\csname bibitem#1\endcsname}%
\let\auto@bib@innerbib\@empty
\bibitem [{\citenamefont {Gross}(1957)}]{Gross1957a}%
  \BibitemOpen
  \bibfield  {author} {\bibinfo {author} {\bibfnamefont {Eugene~P.}\
  \bibnamefont {Gross}},\ }\bibfield  {title} {\enquote {\bibinfo {title}
  {Unified theory of interacting bosons},}\ }\href {\doibase %
  10.1103/PhysRev.106.161} {\bibfield  {journal} {\bibinfo  {journal} {Phys.
  Rev.}\ }\textbf {\bibinfo {volume} {106}},\ \bibinfo {pages} {161} (\bibinfo
  {year} {1957})}\BibitemShut {NoStop}%
\bibitem [{\citenamefont {Andreev}\ and\ \citenamefont
  {Lifshitz}(1969)}]{Andreev1969a}%
  \BibitemOpen
  \bibfield  {author} {\bibinfo {author} {\bibfnamefont {A.~F.}\ \bibnamefont
  {Andreev}}\ and\ \bibinfo {author} {\bibfnamefont {I.~M.}\ \bibnamefont
  {Lifshitz}},\ }\bibfield  {title} {\enquote {\bibinfo {title} {Quantum theory
  of defects in crystals},}\ }\href@noop {} {\bibfield  {journal} {\bibinfo
  {journal} {Sov. Phys. JETP}\ }\textbf {\bibinfo {volume} {29}},\ \bibinfo
  {pages} {1107} (\bibinfo {year} {1969})}\BibitemShut {NoStop}%
\bibitem [{\citenamefont {Chester}(1970)}]{Chester1970a}%
  \BibitemOpen
  \bibfield  {author} {\bibinfo {author} {\bibfnamefont {G.~V.}\ \bibnamefont
  {Chester}},\ }\bibfield  {title} {\enquote {\bibinfo {title} {Speculations on
  {B}ose-{E}instein condensation and quantum crystals},}\ }\href {\doibase %
  10.1103/PhysRevA.2.256} {\bibfield  {journal} {\bibinfo  {journal} {Phys.
  Rev. A}\ }\textbf {\bibinfo {volume} {2}},\ \bibinfo {pages} {256} (\bibinfo
  {year} {1970})}\BibitemShut {NoStop}%
\bibitem [{\citenamefont {Leggett}(1970)}]{Leggett1970a}%
  \BibitemOpen
  \bibfield  {author} {\bibinfo {author} {\bibfnamefont {A.~J.}\ \bibnamefont
  {Leggett}},\ }\bibfield  {title} {\enquote {\bibinfo {title} {Can a solid be
  "superfluid"?}}\ }\href {\doibase 10.1103/PhysRevLett.25.1543} {\bibfield
  {journal} {\bibinfo  {journal} {Phys. Rev. Lett.}\ }\textbf {\bibinfo
  {volume} {25}},\ \bibinfo {pages} {1543} (\bibinfo {year}
  {1970})}\BibitemShut {NoStop}%
\bibitem [{\citenamefont {Boninsegni}\ and\ \citenamefont
  {Prokof'ev}(2012)}]{Boninsegni2012a}%
  \BibitemOpen
  \bibfield  {author} {\bibinfo {author} {\bibfnamefont {Massimo}\ \bibnamefont
  {Boninsegni}}\ and\ \bibinfo {author} {\bibfnamefont {Nikolay~V.}\
  \bibnamefont {Prokof'ev}},\ }\bibfield  {title} {\enquote {\bibinfo {title}
  {\textit{Colloquium} : Supersolids: What and where are they?}}\ }\href
  {\doibase 10.1103/RevModPhys.84.759} {\bibfield  {journal} {\bibinfo
  {journal} {Rev. Mod. Phys.}\ }\textbf {\bibinfo {volume} {84}},\ \bibinfo
  {pages} {759} (\bibinfo {year} {2012})}\BibitemShut {NoStop}%
\bibitem [{\citenamefont {Kim}\ and\ \citenamefont {Chan}(2004)}]{Chan2004a}%
  \BibitemOpen
  \bibfield  {author} {\bibinfo {author} {\bibfnamefont {E.}~\bibnamefont
  {Kim}}\ and\ \bibinfo {author} {\bibfnamefont {M.~H.~W.}\ \bibnamefont
  {Chan}},\ }\bibfield  {title} {\enquote {\bibinfo {title} {Probable
  observation of a supersolid helium phase},}\ }\href {\doibase %
  10.1038/nature02220} {\bibfield  {journal} {\bibinfo  {journal} {Nature}\
  }\textbf {\bibinfo {volume} {427}},\ \bibinfo {pages} {225} (\bibinfo {year}
  {2004})}\BibitemShut {NoStop}%
\bibitem [{\citenamefont {Kim}\ and\ \citenamefont {Chan}(2012)}]{Kim2012a}%
  \BibitemOpen
  \bibfield  {author} {\bibinfo {author} {\bibfnamefont {Duk~Y.}\ \bibnamefont
  {Kim}}\ and\ \bibinfo {author} {\bibfnamefont {Moses H.~W.}\ \bibnamefont
  {Chan}},\ }\bibfield  {title} {\enquote {\bibinfo {title} {Absence of
  supersolidity in solid helium in porous {V}ycor glass},}\ }\href {\doibase %
  10.1103/PhysRevLett.109.155301} {\bibfield  {journal} {\bibinfo  {journal}
  {Phys. Rev. Lett.}\ }\textbf {\bibinfo {volume} {109}},\ \bibinfo {pages}
  {155301} (\bibinfo {year} {2012})}\BibitemShut {NoStop}%
\bibitem [{\citenamefont {Henkel}\ \emph {et~al.}(2010)\citenamefont {Henkel},
  \citenamefont {Nath},\ and\ \citenamefont {Pohl}}]{Henkel2010a}%
  \BibitemOpen
  \bibfield  {author} {\bibinfo {author} {\bibfnamefont {N.}~\bibnamefont
  {Henkel}}, \bibinfo {author} {\bibfnamefont {R.}~\bibnamefont {Nath}}, \ and\
  \bibinfo {author} {\bibfnamefont {T.}~\bibnamefont {Pohl}},\ }\bibfield
  {title} {\enquote {\bibinfo {title} {Three-dimensional roton excitations and
  supersolid formation in {R}ydberg-excited {B}ose-{E}instein condensates},}\
  }\href {\doibase 10.1103/PhysRevLett.104.195302} {\bibfield  {journal}
  {\bibinfo  {journal} {Phys. Rev. Lett.}\ }\textbf {\bibinfo {volume} {104}},\
  \bibinfo {pages} {195302} (\bibinfo {year} {2010})}\BibitemShut {NoStop}%
\bibitem [{\citenamefont {Saccani}\ \emph {et~al.}(2012)\citenamefont
  {Saccani}, \citenamefont {Moroni},\ and\ \citenamefont
  {Boninsegni}}]{Saccani2012a}%
  \BibitemOpen
  \bibfield  {author} {\bibinfo {author} {\bibfnamefont {S.}~\bibnamefont
  {Saccani}}, \bibinfo {author} {\bibfnamefont {S.}~\bibnamefont {Moroni}}, \
  and\ \bibinfo {author} {\bibfnamefont {M.}~\bibnamefont {Boninsegni}},\
  }\bibfield  {title} {\enquote {\bibinfo {title} {Excitation spectrum of a
  supersolid},}\ }\href {\doibase 10.1103/PhysRevLett.108.175301} {\bibfield
  {journal} {\bibinfo  {journal} {Phys. Rev. Lett.}\ }\textbf {\bibinfo
  {volume} {108}},\ \bibinfo {pages} {175301} (\bibinfo {year}
  {2012})}\BibitemShut {NoStop}%
\bibitem [{\citenamefont {Cinti}\ \emph {et~al.}(2010)\citenamefont {Cinti},
  \citenamefont {Jain}, \citenamefont {Boninsegni}, \citenamefont {Micheli},
  \citenamefont {Zoller},\ and\ \citenamefont {Pupillo}}]{Cinti2010a}%
  \BibitemOpen
  \bibfield  {author} {\bibinfo {author} {\bibfnamefont {F.}~\bibnamefont
  {Cinti}}, \bibinfo {author} {\bibfnamefont {P.}~\bibnamefont {Jain}},
  \bibinfo {author} {\bibfnamefont {M.}~\bibnamefont {Boninsegni}}, \bibinfo
  {author} {\bibfnamefont {A.}~\bibnamefont {Micheli}}, \bibinfo {author}
  {\bibfnamefont {P.}~\bibnamefont {Zoller}}, \ and\ \bibinfo {author}
  {\bibfnamefont {G.}~\bibnamefont {Pupillo}},\ }\bibfield  {title} {\enquote
  {\bibinfo {title} {Supersolid droplet crystal in a dipole-blockaded gas},}\
  }\href {\doibase 10.1103/PhysRevLett.105.135301} {\bibfield  {journal}
  {\bibinfo  {journal} {Phys. Rev. Lett.}\ }\textbf {\bibinfo {volume} {105}},\
  \bibinfo {pages} {135301} (\bibinfo {year} {2010})}\BibitemShut {NoStop}%
\bibitem [{\citenamefont {Macr\`{\i}}\ \emph {et~al.}(2013)\citenamefont
  {Macr\`{\i}}, \citenamefont {Maucher}, \citenamefont {Cinti},\ and\
  \citenamefont {Pohl}}]{Macri2013a}%
  \BibitemOpen
  \bibfield  {author} {\bibinfo {author} {\bibfnamefont {T.}~\bibnamefont
  {Macr\`{\i}}}, \bibinfo {author} {\bibfnamefont {F.}~\bibnamefont {Maucher}},
  \bibinfo {author} {\bibfnamefont {F.}~\bibnamefont {Cinti}}, \ and\ \bibinfo
  {author} {\bibfnamefont {T.}~\bibnamefont {Pohl}},\ }\bibfield  {title}
  {\enquote {\bibinfo {title} {Elementary excitations of ultracold soft-core
  bosons across the superfluid-supersolid phase transition},}\ }\href {\doibase %
  10.1103/PhysRevA.87.061602} {\bibfield  {journal} {\bibinfo  {journal} {Phys.
  Rev. A}\ }\textbf {\bibinfo {volume} {87}},\ \bibinfo {pages} {061602}
  (\bibinfo {year} {2013})}\BibitemShut {NoStop}%
\bibitem [{\citenamefont {Ancilotto}\ \emph {et~al.}(2013)\citenamefont
  {Ancilotto}, \citenamefont {Rossi},\ and\ \citenamefont
  {Toigo}}]{Ancilotto2013a}%
  \BibitemOpen
  \bibfield  {author} {\bibinfo {author} {\bibfnamefont {Francesco}\
  \bibnamefont {Ancilotto}}, \bibinfo {author} {\bibfnamefont {Maurizio}\
  \bibnamefont {Rossi}}, \ and\ \bibinfo {author} {\bibfnamefont {Flavio}\
  \bibnamefont {Toigo}},\ }\bibfield  {title} {\enquote {\bibinfo {title}
  {Supersolid structure and excitation spectrum of soft-core bosons in three
  dimensions},}\ }\href {\doibase 10.1103/PhysRevA.88.033618} {\bibfield
  {journal} {\bibinfo  {journal} {Phys. Rev. A}\ }\textbf {\bibinfo {volume}
  {88}},\ \bibinfo {pages} {033618} (\bibinfo {year} {2013})}\BibitemShut
  {NoStop}%
\bibitem [{\citenamefont {Lu}\ \emph {et~al.}(2015)\citenamefont {Lu},
  \citenamefont {Li}, \citenamefont {Petrov},\ and\ \citenamefont
  {Shlyapnikov}}]{Lu2015a}%
  \BibitemOpen
  \bibfield  {author} {\bibinfo {author} {\bibfnamefont {Zhen-Kai}\
  \bibnamefont {Lu}}, \bibinfo {author} {\bibfnamefont {Yun}\ \bibnamefont
  {Li}}, \bibinfo {author} {\bibfnamefont {D.~S.}\ \bibnamefont {Petrov}}, \
  and\ \bibinfo {author} {\bibfnamefont {G.~V.}\ \bibnamefont {Shlyapnikov}},\
  }\bibfield  {title} {\enquote {\bibinfo {title} {Stable dilute supersolid of
  two-dimensional dipolar bosons},}\ }\href {\doibase %
  10.1103/PhysRevLett.115.075303} {\bibfield  {journal} {\bibinfo  {journal}
  {Phys. Rev. Lett.}\ }\textbf {\bibinfo {volume} {115}},\ \bibinfo {pages}
  {075303} (\bibinfo {year} {2015})}\BibitemShut {NoStop}%
\bibitem [{\citenamefont {L{\'e}onard}\ \emph {et~al.}(2017)\citenamefont
  {L{\'e}onard}, \citenamefont {Morales}, \citenamefont {Zupancic},
  \citenamefont {Esslinger},\ and\ \citenamefont {Donner}}]{Leonard2017a}%
  \BibitemOpen
  \bibfield  {author} {\bibinfo {author} {\bibfnamefont {Julian}\ \bibnamefont
  {L{\'e}onard}}, \bibinfo {author} {\bibfnamefont {Andrea}\ \bibnamefont
  {Morales}}, \bibinfo {author} {\bibfnamefont {Philip}\ \bibnamefont
  {Zupancic}}, \bibinfo {author} {\bibfnamefont {Tilman}\ \bibnamefont
  {Esslinger}}, \ and\ \bibinfo {author} {\bibfnamefont {Tobias}\ \bibnamefont
  {Donner}},\ }\bibfield  {title} {\enquote {\bibinfo {title} {Supersolid
  formation in a quantum gas breaking a continuous translational symmetry},}\
  }\href {http://dx.doi.org/10.1038/nature21067} {\bibfield  {journal}
  {\bibinfo  {journal} {Nature}\ }\textbf {\bibinfo {volume} {543}},\ \bibinfo
  {pages} {87} (\bibinfo {year} {2017})}\BibitemShut {NoStop}%
\bibitem [{\citenamefont {Li}\ \emph {et~al.}(2017)\citenamefont {Li},
  \citenamefont {Lee}, \citenamefont {Huang}, \citenamefont {Burchesky},
  \citenamefont {Shteynas}, \citenamefont {Top}, \citenamefont {Jamison},\ and\
  \citenamefont {Ketterle}}]{Li2017a}%
  \BibitemOpen
  \bibfield  {author} {\bibinfo {author} {\bibfnamefont {Jun-Ru}\ \bibnamefont
  {Li}}, \bibinfo {author} {\bibfnamefont {Jeongwon}\ \bibnamefont {Lee}},
  \bibinfo {author} {\bibfnamefont {Wujie}\ \bibnamefont {Huang}}, \bibinfo
  {author} {\bibfnamefont {Sean}\ \bibnamefont {Burchesky}}, \bibinfo {author}
  {\bibfnamefont {Boris}\ \bibnamefont {Shteynas}}, \bibinfo {author}
  {\bibfnamefont {F~{\c{C}}}\ \bibnamefont {Top}}, \bibinfo {author}
  {\bibfnamefont {Alan~O.}\ \bibnamefont {Jamison}}, \ and\ \bibinfo {author}
  {\bibfnamefont {Wolfgang}\ \bibnamefont {Ketterle}},\ }\bibfield  {title}
  {\enquote {\bibinfo {title} {A stripe phase with supersolid properties in
  spin--orbit-coupled {B}ose--{E}instein condensates},}\ }\href
  {http://dx.doi.org/10.1038/nature21431} {\bibfield  {journal} {\bibinfo
  {journal} {Nature}\ }\textbf {\bibinfo {volume} {543}},\ \bibinfo {pages}
  {91} (\bibinfo {year} {2017})}\BibitemShut {NoStop}%
\bibitem [{\citenamefont {Wenzel}\ \emph {et~al.}(2017)\citenamefont {Wenzel},
  \citenamefont {B\"ottcher}, \citenamefont {Langen}, \citenamefont
  {Ferrier-Barbut},\ and\ \citenamefont {Pfau}}]{Wenzel2017a}%
  \BibitemOpen
  \bibfield  {author} {\bibinfo {author} {\bibfnamefont {Matthias}\
  \bibnamefont {Wenzel}}, \bibinfo {author} {\bibfnamefont {Fabian}\
  \bibnamefont {B\"ottcher}}, \bibinfo {author} {\bibfnamefont {Tim}\
  \bibnamefont {Langen}}, \bibinfo {author} {\bibfnamefont {Igor}\ \bibnamefont
  {Ferrier-Barbut}}, \ and\ \bibinfo {author} {\bibfnamefont {Tilman}\
  \bibnamefont {Pfau}},\ }\bibfield  {title} {\enquote {\bibinfo {title}
  {Striped states in a many-body system of tilted dipoles},}\ }\href {\doibase %
  10.1103/PhysRevA.96.053630} {\bibfield  {journal} {\bibinfo  {journal} {Phys.
  Rev. A}\ }\textbf {\bibinfo {volume} {96}},\ \bibinfo {pages} {053630}
  (\bibinfo {year} {2017})}\BibitemShut {NoStop}%
\bibitem [{\citenamefont {Baillie}\ and\ \citenamefont
  {Blakie}(2018)}]{Baillie2018a}%
  \BibitemOpen
  \bibfield  {author} {\bibinfo {author} {\bibfnamefont {D.}~\bibnamefont
  {Baillie}}\ and\ \bibinfo {author} {\bibfnamefont {P.~B.}\ \bibnamefont
  {Blakie}},\ }\bibfield  {title} {\enquote {\bibinfo {title} {Droplet crystal
  ground states of a dipolar {B}ose gas},}\ }\href {\doibase %
  10.1103/PhysRevLett.121.195301} {\bibfield  {journal} {\bibinfo  {journal}
  {Phys. Rev. Lett.}\ }\textbf {\bibinfo {volume} {121}},\ \bibinfo {pages}
  {195301} (\bibinfo {year} {2018})}\BibitemShut {NoStop}%
\bibitem [{\citenamefont {Roccuzzo}\ and\ \citenamefont
  {Ancilotto}(2019)}]{Roccuzzo2019a}%
  \BibitemOpen
  \bibfield  {author} {\bibinfo {author} {\bibfnamefont {Santo~Maria}\
  \bibnamefont {Roccuzzo}}\ and\ \bibinfo {author} {\bibfnamefont {Francesco}\
  \bibnamefont {Ancilotto}},\ }\bibfield  {title} {\enquote {\bibinfo {title}
  {Supersolid behavior of a dipolar {B}ose-{E}instein condensate confined in a
  tube},}\ }\href {\doibase 10.1103/PhysRevA.99.041601} {\bibfield  {journal}
  {\bibinfo  {journal} {Phys. Rev. A}\ }\textbf {\bibinfo {volume} {99}},\
  \bibinfo {pages} {041601} (\bibinfo {year} {2019})}\BibitemShut {NoStop}%
\bibitem [{\citenamefont {Zhang}\ \emph {et~al.}(2019)\citenamefont {Zhang},
  \citenamefont {Maucher},\ and\ \citenamefont {Pohl}}]{Zhang2019a}%
  \BibitemOpen
  \bibfield  {author} {\bibinfo {author} {\bibfnamefont {Yong-Chang}\
  \bibnamefont {Zhang}}, \bibinfo {author} {\bibfnamefont {Fabian}\
  \bibnamefont {Maucher}}, \ and\ \bibinfo {author} {\bibfnamefont {Thomas}\
  \bibnamefont {Pohl}},\ }\bibfield  {title} {\enquote {\bibinfo {title}
  {Supersolidity around a critical point in dipolar {Bose}-{Einstein}
  condensates},}\ }\href {\doibase 10.1103/PhysRevLett.123.015301} {\bibfield
  {journal} {\bibinfo  {journal} {Phys. Rev. Lett.}\ }\textbf {\bibinfo
  {volume} {123}},\ \bibinfo {pages} {015301} (\bibinfo {year}
  {2019})}\BibitemShut {NoStop}%
\bibitem [{\citenamefont {Tanzi}\ \emph
  {et~al.}(2019{\natexlab{a}})\citenamefont {Tanzi}, \citenamefont {Lucioni},
  \citenamefont {Fam\`a}, \citenamefont {Catani}, \citenamefont {Fioretti},
  \citenamefont {Gabbanini}, \citenamefont {Bisset}, \citenamefont {Santos},\
  and\ \citenamefont {Modugno}}]{Tanzi2019a}%
  \BibitemOpen
  \bibfield  {author} {\bibinfo {author} {\bibfnamefont {L.}~\bibnamefont
  {Tanzi}}, \bibinfo {author} {\bibfnamefont {E.}~\bibnamefont {Lucioni}},
  \bibinfo {author} {\bibfnamefont {F.}~\bibnamefont {Fam\`a}}, \bibinfo
  {author} {\bibfnamefont {J.}~\bibnamefont {Catani}}, \bibinfo {author}
  {\bibfnamefont {A.}~\bibnamefont {Fioretti}}, \bibinfo {author}
  {\bibfnamefont {C.}~\bibnamefont {Gabbanini}}, \bibinfo {author}
  {\bibfnamefont {R.~N.}\ \bibnamefont {Bisset}}, \bibinfo {author}
  {\bibfnamefont {L.}~\bibnamefont {Santos}}, \ and\ \bibinfo {author}
  {\bibfnamefont {G.}~\bibnamefont {Modugno}},\ }\bibfield  {title} {\enquote
  {\bibinfo {title} {Observation of a dipolar quantum gas with metastable
  supersolid properties},}\ }\href {\doibase 10.1103/PhysRevLett.122.130405}
  {\bibfield  {journal} {\bibinfo  {journal} {Phys. Rev. Lett.}\ }\textbf
  {\bibinfo {volume} {122}},\ \bibinfo {pages} {130405} (\bibinfo {year}
  {2019}{\natexlab{a}})}\BibitemShut {NoStop}%
\bibitem [{\citenamefont {B\"ottcher}\ \emph {et~al.}(2019)\citenamefont
  {B\"ottcher}, \citenamefont {Schmidt}, \citenamefont {Wenzel}, \citenamefont
  {Hertkorn}, \citenamefont {Guo}, \citenamefont {Langen},\ and\ \citenamefont
  {Pfau}}]{Bottcher2019a}%
  \BibitemOpen
  \bibfield  {author} {\bibinfo {author} {\bibfnamefont {Fabian}\ \bibnamefont
  {B\"ottcher}}, \bibinfo {author} {\bibfnamefont {Jan-Niklas}\ \bibnamefont
  {Schmidt}}, \bibinfo {author} {\bibfnamefont {Matthias}\ \bibnamefont
  {Wenzel}}, \bibinfo {author} {\bibfnamefont {Jens}\ \bibnamefont {Hertkorn}},
  \bibinfo {author} {\bibfnamefont {Mingyang}\ \bibnamefont {Guo}}, \bibinfo
  {author} {\bibfnamefont {Tim}\ \bibnamefont {Langen}}, \ and\ \bibinfo
  {author} {\bibfnamefont {Tilman}\ \bibnamefont {Pfau}},\ }\bibfield  {title}
  {\enquote {\bibinfo {title} {Transient supersolid properties in an array of
  dipolar quantum droplets},}\ }\href {\doibase 10.1103/PhysRevX.9.011051}
  {\bibfield  {journal} {\bibinfo  {journal} {Phys. Rev. X}\ }\textbf {\bibinfo
  {volume} {9}},\ \bibinfo {pages} {011051} (\bibinfo {year}
  {2019})}\BibitemShut {NoStop}%
\bibitem [{\citenamefont {Chomaz}\ \emph {et~al.}(2019)\citenamefont {Chomaz},
  \citenamefont {Petter}, \citenamefont {Ilzh\"ofer}, \citenamefont {Natale},
  \citenamefont {Trautmann}, \citenamefont {Politi}, \citenamefont
  {Durastante}, \citenamefont {van Bijnen}, \citenamefont {Patscheider},
  \citenamefont {Sohmen}, \citenamefont {Mark},\ and\ \citenamefont
  {Ferlaino}}]{Chomaz2019a}%
  \BibitemOpen
  \bibfield  {author} {\bibinfo {author} {\bibfnamefont {L.}~\bibnamefont
  {Chomaz}}, \bibinfo {author} {\bibfnamefont {D.}~\bibnamefont {Petter}},
  \bibinfo {author} {\bibfnamefont {P.}~\bibnamefont {Ilzh\"ofer}}, \bibinfo
  {author} {\bibfnamefont {G.}~\bibnamefont {Natale}}, \bibinfo {author}
  {\bibfnamefont {A.}~\bibnamefont {Trautmann}}, \bibinfo {author}
  {\bibfnamefont {C.}~\bibnamefont {Politi}}, \bibinfo {author} {\bibfnamefont
  {G.}~\bibnamefont {Durastante}}, \bibinfo {author} {\bibfnamefont {R.~M.~W.}\
  \bibnamefont {van Bijnen}}, \bibinfo {author} {\bibfnamefont
  {A.}~\bibnamefont {Patscheider}}, \bibinfo {author} {\bibfnamefont
  {M.}~\bibnamefont {Sohmen}}, \bibinfo {author} {\bibfnamefont {M.~J.}\
  \bibnamefont {Mark}}, \ and\ \bibinfo {author} {\bibfnamefont
  {F.}~\bibnamefont {Ferlaino}},\ }\bibfield  {title} {\enquote {\bibinfo
  {title} {Long-lived and transient supersolid behaviors in dipolar quantum
  gases},}\ }\href {\doibase 10.1103/PhysRevX.9.021012} {\bibfield  {journal}
  {\bibinfo  {journal} {Phys. Rev. X}\ }\textbf {\bibinfo {volume} {9}},\
  \bibinfo {pages} {021012} (\bibinfo {year} {2019})}\BibitemShut {NoStop}%
\bibitem [{\citenamefont {Tanzi}\ \emph
  {et~al.}(2019{\natexlab{b}})\citenamefont {Tanzi}, \citenamefont {Roccuzzo},
  \citenamefont {Lucioni}, \citenamefont {Fam{\`a}}, \citenamefont {Fioretti},
  \citenamefont {Gabbanini}, \citenamefont {Modugno}, \citenamefont {Recati},\
  and\ \citenamefont {Stringari}}]{Tanzi2019b}%
  \BibitemOpen
  \bibfield  {author} {\bibinfo {author} {\bibfnamefont {L.}~\bibnamefont
  {Tanzi}}, \bibinfo {author} {\bibfnamefont {S.~M.}\ \bibnamefont {Roccuzzo}},
  \bibinfo {author} {\bibfnamefont {E.}~\bibnamefont {Lucioni}}, \bibinfo
  {author} {\bibfnamefont {F.}~\bibnamefont {Fam{\`a}}}, \bibinfo {author}
  {\bibfnamefont {A.}~\bibnamefont {Fioretti}}, \bibinfo {author}
  {\bibfnamefont {C.}~\bibnamefont {Gabbanini}}, \bibinfo {author}
  {\bibfnamefont {G.}~\bibnamefont {Modugno}}, \bibinfo {author} {\bibfnamefont
  {A.}~\bibnamefont {Recati}}, \ and\ \bibinfo {author} {\bibfnamefont
  {S.}~\bibnamefont {Stringari}},\ }\bibfield  {title} {\enquote {\bibinfo
  {title} {Supersolid symmetry breaking from compressional oscillations in a
  dipolar quantum gas},}\ }\href {\doibase 10.1038/s41586-019-1568-6}
  {\bibfield  {journal} {\bibinfo  {journal} {Nature}\ }\textbf {\bibinfo
  {volume} {574}},\ \bibinfo {pages} {382} (\bibinfo {year}
  {2019}{\natexlab{b}})}\BibitemShut {NoStop}%
\bibitem [{\citenamefont {Guo}\ \emph {et~al.}(2019)\citenamefont {Guo},
  \citenamefont {B{\"o}ttcher}, \citenamefont {Hertkorn}, \citenamefont
  {Schmidt}, \citenamefont {Wenzel}, \citenamefont {B{\"u}chler}, \citenamefont
  {Langen},\ and\ \citenamefont {Pfau}}]{Guo2019a}%
  \BibitemOpen
  \bibfield  {author} {\bibinfo {author} {\bibfnamefont {Mingyang}\
  \bibnamefont {Guo}}, \bibinfo {author} {\bibfnamefont {Fabian}\ \bibnamefont
  {B{\"o}ttcher}}, \bibinfo {author} {\bibfnamefont {Jens}\ \bibnamefont
  {Hertkorn}}, \bibinfo {author} {\bibfnamefont {Jan-Niklas}\ \bibnamefont
  {Schmidt}}, \bibinfo {author} {\bibfnamefont {Matthias}\ \bibnamefont
  {Wenzel}}, \bibinfo {author} {\bibfnamefont {Hans~Peter}\ \bibnamefont
  {B{\"u}chler}}, \bibinfo {author} {\bibfnamefont {Tim}\ \bibnamefont
  {Langen}}, \ and\ \bibinfo {author} {\bibfnamefont {Tilman}\ \bibnamefont
  {Pfau}},\ }\bibfield  {title} {\enquote {\bibinfo {title} {The low-energy
  goldstone mode in a trapped dipolar supersolid},}\ }\href {\doibase %
  10.1038/s41586-019-1569-5} {\bibfield  {journal} {\bibinfo  {journal}
  {Nature}\ }\textbf {\bibinfo {volume} {564}},\ \bibinfo {pages} {386}
  (\bibinfo {year} {2019})}\BibitemShut {NoStop}%
\bibitem [{\citenamefont {Natale}\ \emph {et~al.}(2019)\citenamefont {Natale},
  \citenamefont {van Bijnen}, \citenamefont {Patscheider}, \citenamefont
  {Petter}, \citenamefont {Mark}, \citenamefont {Chomaz},\ and\ \citenamefont
  {Ferlaino}}]{Natale2019a}%
  \BibitemOpen
  \bibfield  {author} {\bibinfo {author} {\bibfnamefont {G.}~\bibnamefont
  {Natale}}, \bibinfo {author} {\bibfnamefont {R.~M.~W.}\ \bibnamefont {van
  Bijnen}}, \bibinfo {author} {\bibfnamefont {A.}~\bibnamefont {Patscheider}},
  \bibinfo {author} {\bibfnamefont {D.}~\bibnamefont {Petter}}, \bibinfo
  {author} {\bibfnamefont {M.~J.}\ \bibnamefont {Mark}}, \bibinfo {author}
  {\bibfnamefont {L.}~\bibnamefont {Chomaz}}, \ and\ \bibinfo {author}
  {\bibfnamefont {F.}~\bibnamefont {Ferlaino}},\ }\bibfield  {title} {\enquote
  {\bibinfo {title} {Excitation spectrum of a trapped dipolar supersolid and
  its experimental evidence},}\ }\href {\doibase %
  10.1103/PhysRevLett.123.050402} {\bibfield  {journal} {\bibinfo  {journal}
  {Phys. Rev. Lett.}\ }\textbf {\bibinfo {volume} {123}},\ \bibinfo {pages}
  {050402} (\bibinfo {year} {2019})}\BibitemShut {NoStop}%
\bibitem [{\citenamefont {Ferrier-Barbut}\ \emph {et~al.}(2016)\citenamefont
  {Ferrier-Barbut}, \citenamefont {Kadau}, \citenamefont {Schmitt},
  \citenamefont {Wenzel},\ and\ \citenamefont {Pfau}}]{Ferrier-Barbut2016a}%
  \BibitemOpen
  \bibfield  {author} {\bibinfo {author} {\bibfnamefont {Igor}\ \bibnamefont
  {Ferrier-Barbut}}, \bibinfo {author} {\bibfnamefont {Holger}\ \bibnamefont
  {Kadau}}, \bibinfo {author} {\bibfnamefont {Matthias}\ \bibnamefont
  {Schmitt}}, \bibinfo {author} {\bibfnamefont {Matthias}\ \bibnamefont
  {Wenzel}}, \ and\ \bibinfo {author} {\bibfnamefont {Tilman}\ \bibnamefont
  {Pfau}},\ }\bibfield  {title} {\enquote {\bibinfo {title} {Observation of
  quantum droplets in a strongly dipolar {B}ose gas},}\ }\href {\doibase %
  10.1103/PhysRevLett.116.215301} {\bibfield  {journal} {\bibinfo  {journal}
  {Phys. Rev. Lett.}\ }\textbf {\bibinfo {volume} {116}},\ \bibinfo {pages}
  {215301} (\bibinfo {year} {2016})}\BibitemShut {NoStop}%
\bibitem [{\citenamefont {Chomaz}\ \emph {et~al.}(2016)\citenamefont {Chomaz},
  \citenamefont {Baier}, \citenamefont {Petter}, \citenamefont {Mark},
  \citenamefont {W\"achtler}, \citenamefont {Santos},\ and\ \citenamefont
  {Ferlaino}}]{Chomaz2016a}%
  \BibitemOpen
  \bibfield  {author} {\bibinfo {author} {\bibfnamefont {L.}~\bibnamefont
  {Chomaz}}, \bibinfo {author} {\bibfnamefont {S.}~\bibnamefont {Baier}},
  \bibinfo {author} {\bibfnamefont {D.}~\bibnamefont {Petter}}, \bibinfo
  {author} {\bibfnamefont {M.~J.}\ \bibnamefont {Mark}}, \bibinfo {author}
  {\bibfnamefont {F.}~\bibnamefont {W\"achtler}}, \bibinfo {author}
  {\bibfnamefont {L.}~\bibnamefont {Santos}}, \ and\ \bibinfo {author}
  {\bibfnamefont {F.}~\bibnamefont {Ferlaino}},\ }\bibfield  {title} {\enquote
  {\bibinfo {title} {Quantum-fluctuation-driven crossover from a dilute
  {B}ose-{E}instein condensate to a macrodroplet in a dipolar quantum fluid},}\
  }\href {\doibase 10.1103/PhysRevX.6.041039} {\bibfield  {journal} {\bibinfo
  {journal} {Phys. Rev. X}\ }\textbf {\bibinfo {volume} {6}},\ \bibinfo {pages}
  {041039} (\bibinfo {year} {2016})}\BibitemShut {NoStop}%
\bibitem [{\citenamefont {W\"achtler}\ and\ \citenamefont
  {Santos}(2016)}]{Wachtler2016a}%
  \BibitemOpen
  \bibfield  {author} {\bibinfo {author} {\bibfnamefont {F.}~\bibnamefont
  {W\"achtler}}\ and\ \bibinfo {author} {\bibfnamefont {L.}~\bibnamefont
  {Santos}},\ }\bibfield  {title} {\enquote {\bibinfo {title} {Quantum
  filaments in dipolar {B}ose-{E}instein condensates},}\ }\href {\doibase %
  10.1103/PhysRevA.93.061603} {\bibfield  {journal} {\bibinfo  {journal} {Phys.
  Rev. A}\ }\textbf {\bibinfo {volume} {93}},\ \bibinfo {pages} {061603(R)}
  (\bibinfo {year} {2016})}\BibitemShut {NoStop}%
\bibitem [{\citenamefont {Bisset}\ \emph {et~al.}(2016)\citenamefont {Bisset},
  \citenamefont {Wilson}, \citenamefont {Baillie},\ and\ \citenamefont
  {Blakie}}]{Bisset2016a}%
  \BibitemOpen
  \bibfield  {author} {\bibinfo {author} {\bibfnamefont {R.~N.}\ \bibnamefont
  {Bisset}}, \bibinfo {author} {\bibfnamefont {R.~M.}\ \bibnamefont {Wilson}},
  \bibinfo {author} {\bibfnamefont {D.}~\bibnamefont {Baillie}}, \ and\
  \bibinfo {author} {\bibfnamefont {P.~B.}\ \bibnamefont {Blakie}},\ }\bibfield
   {title} {\enquote {\bibinfo {title} {Ground-state phase diagram of a dipolar
  condensate with quantum fluctuations},}\ }\href {\doibase %
  10.1103/PhysRevA.94.033619} {\bibfield  {journal} {\bibinfo  {journal} {Phys.
  Rev. A}\ }\textbf {\bibinfo {volume} {94}},\ \bibinfo {pages} {033619}
  (\bibinfo {year} {2016})}\BibitemShut {NoStop}%
\bibitem [{\citenamefont {Sep\'ulveda}\ \emph {et~al.}(2008)\citenamefont
  {Sep\'ulveda}, \citenamefont {Josserand},\ and\ \citenamefont
  {Rica}}]{Sepulveda2008a}%
  \BibitemOpen
  \bibfield  {author} {\bibinfo {author} {\bibfnamefont {N\'estor}\
  \bibnamefont {Sep\'ulveda}}, \bibinfo {author} {\bibfnamefont {Christophe}\
  \bibnamefont {Josserand}}, \ and\ \bibinfo {author} {\bibfnamefont {Sergio}\
  \bibnamefont {Rica}},\ }\bibfield  {title} {\enquote {\bibinfo {title}
  {Nonclassical rotational inertia fraction in a one-dimensional model of a
  supersolid},}\ }\href {\doibase 10.1103/PhysRevB.77.054513} {\bibfield
  {journal} {\bibinfo  {journal} {Phys. Rev. B}\ }\textbf {\bibinfo {volume}
  {77}},\ \bibinfo {pages} {054513} (\bibinfo {year} {2008})}\BibitemShut
  {NoStop}%
\bibitem [{\citenamefont {Blakie}\ \emph {et~al.}(2020)\citenamefont {Blakie},
  \citenamefont {Baillie},\ and\ \citenamefont {Pal}}]{Blakie2020a}%
  \BibitemOpen
  \bibfield  {author} {\bibinfo {author} {\bibfnamefont {P.~Blair}\
  \bibnamefont {Blakie}}, \bibinfo {author} {\bibfnamefont {D.}~\bibnamefont
  {Baillie}}, \ and\ \bibinfo {author} {\bibfnamefont {Sukla}\ \bibnamefont
  {Pal}},\ }\href@noop {} {\enquote {\bibinfo {title} {Variational theory for
  the ground state and collective excitations of an elongated dipolar
  condensate},}\ } \Eprint
  {http://arxiv.org/abs/2004.09859} {arXiv:2004.09859}
  \BibitemShut {NoStop}%
\bibitem [{\citenamefont {Lima}\ and\ \citenamefont
  {Pelster}(2011)}]{Lima2011a}%
  \BibitemOpen
  \bibfield  {author} {\bibinfo {author} {\bibfnamefont {Aristeu R.~P.}\
  \bibnamefont {Lima}}\ and\ \bibinfo {author} {\bibfnamefont {Axel}\
  \bibnamefont {Pelster}},\ }\bibfield  {title} {\enquote {\bibinfo {title}
  {Quantum fluctuations in dipolar {B}ose gases},}\ }\href {\doibase %
  10.1103/PhysRevA.84.041604} {\bibfield  {journal} {\bibinfo  {journal} {Phys.
  Rev. A}\ }\textbf {\bibinfo {volume} {84}},\ \bibinfo {pages} {041604}
  (\bibinfo {year} {2011})}\BibitemShut {NoStop}%
\bibitem [{\citenamefont {Sep{\'u}lveda}\ \emph {et~al.}(2010)\citenamefont
  {Sep{\'u}lveda}, \citenamefont {Josserand},\ and\ \citenamefont
  {Rica}}]{Sepulveda2010a}%
  \BibitemOpen
  \bibfield  {author} {\bibinfo {author} {\bibfnamefont {N.}~\bibnamefont
  {Sep{\'u}lveda}}, \bibinfo {author} {\bibfnamefont {C.}~\bibnamefont
  {Josserand}}, \ and\ \bibinfo {author} {\bibfnamefont {S.}~\bibnamefont
  {Rica}},\ }\bibfield  {title} {\enquote {\bibinfo {title} {Superfluid density
  in a two-dimensional model of supersolid},}\ }\href {\doibase %
  10.1140/epjb/e2010-10176-y} {\bibfield  {journal} {\bibinfo  {journal} {Euro.
  Phys. J. B}\ }\textbf {\bibinfo {volume} {78}},\ \bibinfo {pages} {439}
  (\bibinfo {year} {2010})}\BibitemShut {NoStop}%
\bibitem [{\citenamefont {Santos}\ \emph {et~al.}(2003)\citenamefont {Santos},
  \citenamefont {Shlyapnikov},\ and\ \citenamefont {Lewenstein}}]{Santos2003a}%
  \BibitemOpen
  \bibfield  {author} {\bibinfo {author} {\bibfnamefont {L.}~\bibnamefont
  {Santos}}, \bibinfo {author} {\bibfnamefont {G.~V.}\ \bibnamefont
  {Shlyapnikov}}, \ and\ \bibinfo {author} {\bibfnamefont {M.}~\bibnamefont
  {Lewenstein}},\ }\bibfield  {title} {\enquote {\bibinfo {title} {Roton-maxon
  spectrum and stability of trapped dipolar {B}ose-{E}instein condensates},}\
  }\href {\doibase 10.1103/PhysRevLett.90.250403} {\bibfield  {journal}
  {\bibinfo  {journal} {Phys. Rev. Lett.}\ }\textbf {\bibinfo {volume} {90}},\
  \bibinfo {pages} {250403} (\bibinfo {year} {2003})}\BibitemShut {NoStop}%
\bibitem [{\citenamefont {Chomaz}\ \emph {et~al.}(2018)\citenamefont {Chomaz},
  \citenamefont {van Bijnen}, \citenamefont {Petter}, \citenamefont {Faraoni},
  \citenamefont {Baier}, \citenamefont {Becher}, \citenamefont {Mark},
  \citenamefont {W{\"a}chtler}, \citenamefont {Santos},\ and\ \citenamefont
  {Ferlaino}}]{Chomaz2018a}%
  \BibitemOpen
  \bibfield  {author} {\bibinfo {author} {\bibfnamefont {L.}~\bibnamefont
  {Chomaz}}, \bibinfo {author} {\bibfnamefont {R.~M.~W.}\ \bibnamefont {van
  Bijnen}}, \bibinfo {author} {\bibfnamefont {D.}~\bibnamefont {Petter}},
  \bibinfo {author} {\bibfnamefont {G.}~\bibnamefont {Faraoni}}, \bibinfo
  {author} {\bibfnamefont {S.}~\bibnamefont {Baier}}, \bibinfo {author}
  {\bibfnamefont {J.~H.}\ \bibnamefont {Becher}}, \bibinfo {author}
  {\bibfnamefont {M.~J.}\ \bibnamefont {Mark}}, \bibinfo {author}
  {\bibfnamefont {F.}~\bibnamefont {W{\"a}chtler}}, \bibinfo {author}
  {\bibfnamefont {L.}~\bibnamefont {Santos}}, \ and\ \bibinfo {author}
  {\bibfnamefont {F.}~\bibnamefont {Ferlaino}},\ }\bibfield  {title} {\enquote
  {\bibinfo {title} {Observation of roton mode population in a dipolar quantum
  gas},}\ }\href {\doibase 10.1038/s41567-018-0054-7} {\bibfield  {journal}
  {\bibinfo  {journal} {Nat. Phys.}\ }\textbf {\bibinfo {volume} {14}},\
  \bibinfo {pages} {442} (\bibinfo {year} {2018})}\BibitemShut {NoStop}%
\bibitem [{\citenamefont {Lima}\ and\ \citenamefont
  {Pelster}(2010)}]{Lima2010a}%
  \BibitemOpen
  \bibfield  {author} {\bibinfo {author} {\bibfnamefont {Aristeu R.~P.}\
  \bibnamefont {Lima}}\ and\ \bibinfo {author} {\bibfnamefont {Axel}\
  \bibnamefont {Pelster}},\ }\bibfield  {title} {\enquote {\bibinfo {title}
  {Dipolar {F}ermi gases in anisotropic traps},}\ }\href {\doibase %
  10.1103/PhysRevA.81.063629} {\bibfield  {journal} {\bibinfo  {journal} {Phys.
  Rev. A}\ }\textbf {\bibinfo {volume} {81}},\ \bibinfo {pages} {063629}
  (\bibinfo {year} {2010})}\BibitemShut {NoStop}%
\bibitem [{\citenamefont {Blakie}(2016)}]{Blakie2016a}%
  \BibitemOpen
  \bibfield  {author} {\bibinfo {author} {\bibfnamefont {P.~B.}\ \bibnamefont
  {Blakie}},\ }\bibfield  {title} {\enquote {\bibinfo {title} {Properties of a
  dipolar condensate with three-body interactions},}\ }\href {\doibase %
  10.1103/PhysRevA.93.033644} {\bibfield  {journal} {\bibinfo  {journal} {Phys.
  Rev. A}\ }\textbf {\bibinfo {volume} {93}},\ \bibinfo {pages} {033644}
  (\bibinfo {year} {2016})}\BibitemShut {NoStop}%
\bibitem [{\citenamefont {Ferrier-Barbut}\ \emph {et~al.}(2018)\citenamefont
  {Ferrier-Barbut}, \citenamefont {Wenzel}, \citenamefont {Schmitt},
  \citenamefont {B\"ottcher},\ and\ \citenamefont
  {Pfau}}]{Ferrier-Barbut2018a}%
  \BibitemOpen
  \bibfield  {author} {\bibinfo {author} {\bibfnamefont {Igor}\ \bibnamefont
  {Ferrier-Barbut}}, \bibinfo {author} {\bibfnamefont {Matthias}\ \bibnamefont
  {Wenzel}}, \bibinfo {author} {\bibfnamefont {Matthias}\ \bibnamefont
  {Schmitt}}, \bibinfo {author} {\bibfnamefont {Fabian}\ \bibnamefont
  {B\"ottcher}}, \ and\ \bibinfo {author} {\bibfnamefont {Tilman}\ \bibnamefont
  {Pfau}},\ }\bibfield  {title} {\enquote {\bibinfo {title} {Onset of a
  modulational instability in trapped dipolar {B}ose-{E}instein condensates},}\
  }\href {\doibase 10.1103/PhysRevA.97.011604} {\bibfield  {journal} {\bibinfo
  {journal} {Phys. Rev. A}\ }\textbf {\bibinfo {volume} {97}},\ \bibinfo
  {pages} {011604(R)} (\bibinfo {year} {2018})}\BibitemShut {NoStop}%
\end{thebibliography}
%

\end{document}